\documentclass[conference]{IEEEtran}
\IEEEoverridecommandlockouts

\usepackage{cite}
\usepackage{amsmath,amssymb,amsfonts}
\usepackage{algorithmic}
\usepackage{graphicx}
\usepackage{textcomp}
\usepackage{xcolor}
\usepackage[T1]{fontenc}
\usepackage{siunitx}
\usepackage{gensymb}
\usepackage{booktabs}
\newcommand{\hide}[1]{}
\newcommand{\bdmath}{\begin{dmath}}
	\newcommand{\edmath}{\end{dmath}}
\newcommand{\beq}{\begin{equation}}
	\newcommand{\eeq}{\end{equation}}
\newcommand{\bdm}{\begin{displaymath}}
	\newcommand{\edm}{\end{displaymath}}
\newcommand{\bea}{\begin{eqnarray}}
	\newcommand{\eea}{\end{eqnarray}}
\newcommand{\beal}{\beq \begin{array}{ll}}
	\newcommand{\eeal}{\end{array} \eeq}
\newcommand{\beas}{\begin{eqnarray*}}
	\newcommand{\eeas}{\end{eqnarray*}}
	\newcommand{\ea}{\end{array}}
\newcommand{\bit}{\begin{itemize}}
	\newcommand{\eit}{\end{itemize}}
\newcommand{\ben}{\begin{enumerate}}
	\newcommand{\een}{\end{enumerate}}

\makeatletter %
\newcommand{\linebreakand}{%
  \end{@IEEEauthorhalign}
  \hfill\mbox{}\par
  \mbox{}\hfill\begin{@IEEEauthorhalign}
}
\makeatother %

\def\BibTeX{{\rm B\kern-.05em{\sc i\kern-.025em b}\kern-.08em
    T\kern-.1667em\lower.7ex\hbox{E}\kern-.125emX}}
    
\begin{document}
\begin{table*}
\centering
\normalsize
This work has been submitted to the IEEE for possible publication. \\Copyright 
may be transferred without notice, after which this version may no longer be 
accessible.
\end{table*}
\newpage

\title{
3D Freehand Ultrasound using Visual Inertial and Deep Inertial Odometry for Measuring Patellar Tracking
\thanks{This work was supported by the EPSRC ORCA Robotics Hub (EP/R026173/1), the EU H2020 Project THING (Grant ID 780883), a Royal Society University Research Fellowship (Fallon), Versus Arthritis MedTech PoC Grant (22702) (Mellon \& Tu) and Orthopaedic Research UK Early-career Research Fellowship (Tu)}
}

\author{\IEEEauthorblockN{Russell Buchanan}
\IEEEauthorblockA{
\textit{Institute of Perception, Action and Behaviour} \\
\textit{University of Edinburgh}\\
Edinburgh, UK \\
russell.buchanan@ed.ac.uk}
\and
\IEEEauthorblockN{S Jack Tu}
\IEEEauthorblockA{
\textit{NDORMS} \\
\textit{University of Oxford}\\
Oxford, UK \\
jack.tu@ndorms.ox.ac.uk}
\and
\IEEEauthorblockN{Marco Camurri}
\IEEEauthorblockA{
\textit{Faculty of Engineering} \\
\textit{Free University of Bozen-Bolzano}\\
Bolzano, Italy \\
marco.camurri@unibz.it}
\and

\linebreakand %

\IEEEauthorblockN{Stephen J Mellon}
\IEEEauthorblockA{
\textit{NDORMS} \\
\textit{University of Oxford}\\
Oxford, UK \\
stephen.mellon@ndorms.ox.ac.uk}
\and
\IEEEauthorblockN{Maurice Fallon}
\IEEEauthorblockA{\textit{Oxford Robotics Institute} \\
\textit{University of Oxford}\\
Oxford, UK\\
mfallon@robots.ox.ac.uk}
}

\maketitle
\begin{abstract}
Patellofemoral joint (PFJ) issues affect one in four people, with 20\% experiencing chronic knee pain despite treatment. Poor outcomes and pain after knee replacement surgery are often linked to patellar mal-tracking. Traditional imaging methods like CT and MRI face challenges, including cost and metal artefacts, and there's currently no ideal way to observe joint motion without issues such as soft tissue artefacts or radiation exposure. A new system to monitor joint motion could significantly improve understanding of PFJ dynamics, aiding in better patient care and outcomes.
\par Combining 2D ultrasound with motion tracking for 3D reconstruction of the joint using semantic segmentation and position registration can be a solution. However, the need for expensive external infrastructure to estimate the trajectories of the scanner remains the main limitation to implementing 3D bone reconstruction from handheld ultrasound scanning clinically.
\par We proposed the Visual-Inertial Odometry (VIO) and the deep learning-based inertial-only odometry methods as alternatives to motion capture for tracking a handheld ultrasound scanner. The 3D reconstruction generated by these methods has demonstrated potential for assessing the PFJ and for further measurements from free-hand ultrasound scans. The results show that the VIO method performs as well as the motion capture method, with average reconstruction errors of  \SI{1.25}{\milli\meter} and \SI{1.21}{\milli\meter}, respectively. The VIO method is the first infrastructure-free method for 3D reconstruction of bone from wireless handheld ultrasound scanning with an accuracy comparable to methods that require external infrastructure.
\end{abstract}

\begin{IEEEkeywords}
3D freehand ultrasound, Deep Learning, Motion estimation, Inertial measurement unit
\end{IEEEkeywords}

\section{Introduction}
In the UK, approximately 100,000 people undergo total knee arthroplasty (TKA) each year. However, around 20\% of TKA patients experience dissatisfaction afterwards\cite{petersen2015,van2017}, typically due to ongoing pain. Abnormal patellofemoral joint (PFJ) kinematics can cause pain after TKA, specifically patellar maltracking \cite{belvedere2014tibio,pulavarti2014}. There are several surgical interventions to address patellar maltracking, but their outcomes vary, and they are only applied to patients with obvious patellar maltracking, such as subluxation and dislocation. A method for tracking the patella during knee flexion in patients with painful TKA could provide clinical insight and help identify less obvious forms of patellar maltracking. This, in turn, could be addressed by an appropriate intervention.

Ultrasound has been shown to have enormous potential for bone visualisation and is a low-cost and radiation-free alternative to MRI and X-ray imaging \cite{Nazarian2008}. Consequently, there has been significant interest in using ultrasound for 3D reconstruction of bone. The most accurate 3D freehand ultrasound approaches require external infrastructure to facilitate device tracking, such as a motion capture system~\cite{Swiatek2014, Jia2017} or electromagnetic markers~\cite{Mahfouz2021} for US probe localisation; or the use of a robot arm for scanning~\cite{Kerr2017}. Unfortunately, these methods require significant infrastructure which increases costs and reduces the mobility of a potential device and could hinder clinical implementation.

\begin{figure*}
	\centering
	\includegraphics[width=0.95\linewidth]{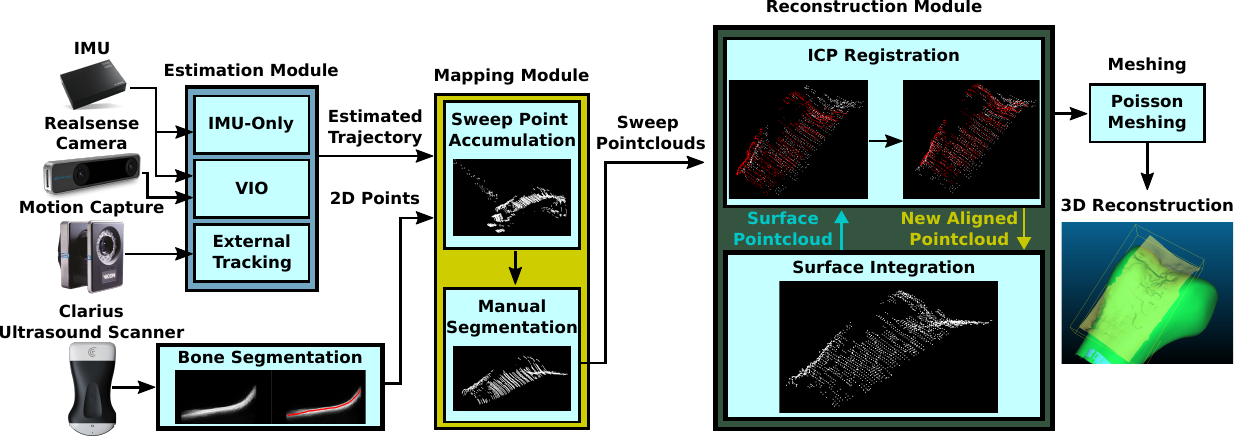}
	\caption{Flow diagram of complete reconstruction pipeline. The estimation module computes the scanning trajectory using either IMU-only, VIO or motion capture external tracking. Separately, 2D ultrasound points are segmented for the presence of bone. These points are projected in 3D using the estimated trajectory and accumulated into separate point clouds for each sweep over the bone. After manually segmenting out the surface, the point clouds are aligned using ICP and integrated into the surface estimate. Finally, once all the sweeps have been integrated, the final surface is sampled and converted to a mesh using the Poisson method.}
	\label{fig:block-diagram}
 \vspace{-4mm}
\end{figure*}

Freehand 3D ultrasound can be achieved without external infrastructure by measuring the elevational speckle decorrelation between frames\cite{GEE2006}. However, it cannot fully capture the complexity of ultrasound image formation. Reference \cite{Prevost2018} presented a method for creating 3D reconstructions using a convolutions neural network (CNN) for motion estimation between ultrasound frames. However, these image-based methods tend to have a significant reconstruction error in at least one plane that is difficult to remove without external localisation. 

``State estimation'' is a well-established research area for tracking the position and orientation of robots without external infrastructures in robotics \cite{Barfoot2017}. Visual Odometry (VO) is one of the research areas within this field, which involves tracking visual features in the environment using a camera to estimate motion\cite{MurArtal2017}. This approach is often combined with Inertial Measurement Units (IMUs), which are small, low-cost sensors capable of measuring acceleration and rotation rates. VIO has superior robustness to visual challenges, such as darkness or motion blur.

Recently, there have been several works applying deep learning to inertial data to enable low-drift, IMU-only odometry \cite{Chen2018, Liu2020, Buchanan2021CORL}. These methods train neural networks to learn motion models of, for example, pedestrians or robots. The network can then predict displacement or velocity from inertial data.

In this study, we used robotic state estimation techniques to track a handheld ultrasound scanner. Our objective was to explore the feasibility of infrastructure-free methods for generating a 3D reconstruction of the bone surface from ultrasound imaging. We also developed a motion model of the scanner by training a neural network with only IMU sensing. Finally, we compared the effectiveness of these two methods against a baseline established by external motion capture tracking of the scanner.

\section{Methods}
We evaluated three different methods for tracking the ultrasound scanner: motion capture, visual-inertial odometry (VIO), and deep learning IMU-only odometry. An OptiTrack system was used to capture the motion of the ultrasound scanner with reflective markers. For VIO, we used a factor graph-based approach \cite{Buchanan2021CORL} to combine data from a stereo camera and an IMU mounted on the scanner. To implement the IMU-only method, we trained a neural network to predict position increments based on IMU data and combined it with classical IMU integration, which was similar to previous work \cite{Wisth2022}. The full system pipeline is shown in Fig. \ref{fig:block-diagram}

\subsection{Experimental Setup}
\label{sec:experiment-setup}
\begin{figure}
\centering
\includegraphics[width=0.9\columnwidth]{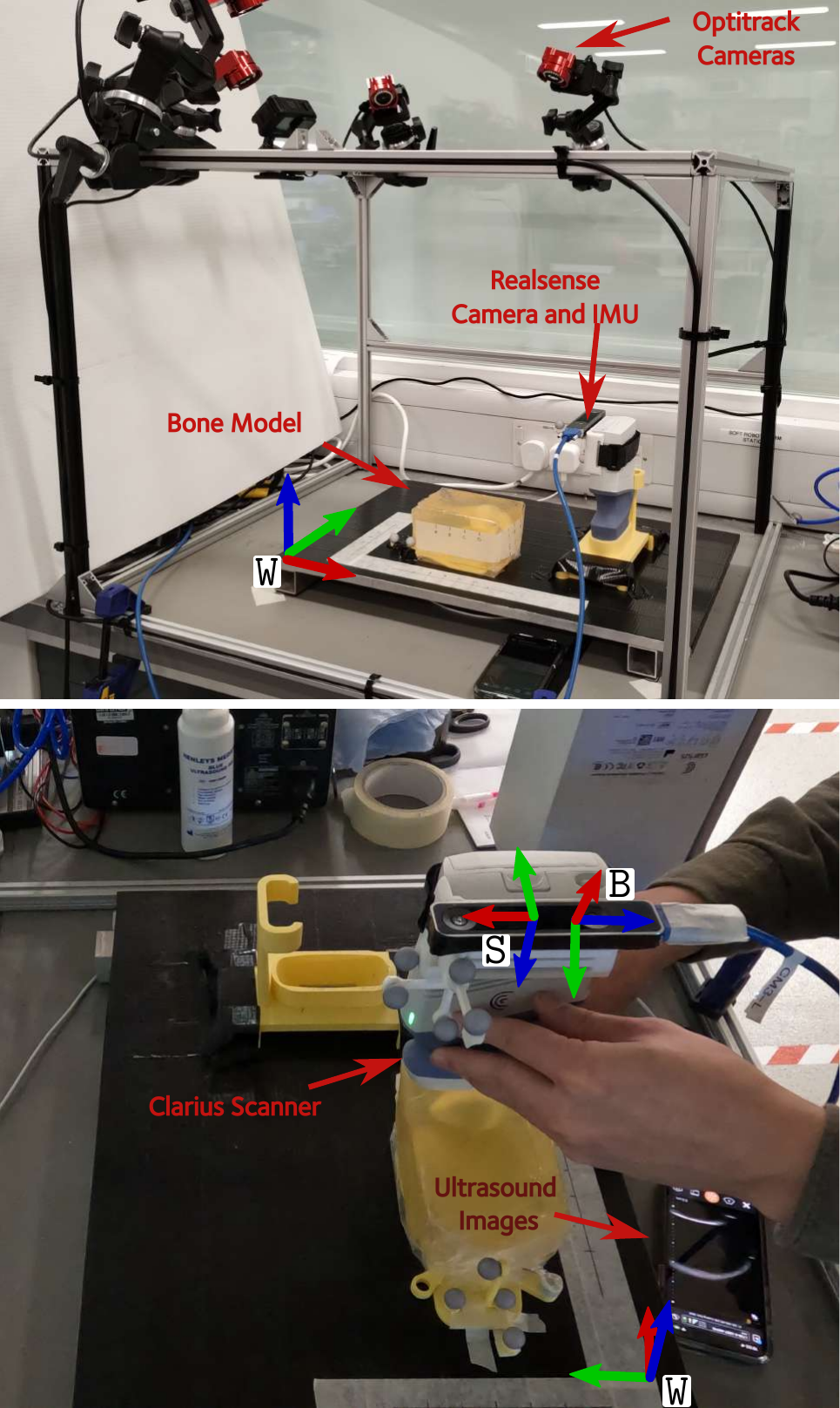}
\caption{Experimental setup for collecting ultrasound dataset. A Realsense
stereo camera is mounted on the Clarius ultrasound scanner and a mount
holds the scanner in place. The human operator picks up the scanner and
passes it over the bone model several times before returning it to the mount.}
\label{fig:us-experiment-setup}
\end{figure}

The experimental setup used is shown in Fig. \ref{fig:us-experiment-setup}. An Intel Realsense T265 stereo camera, which includes a Bosch BMI085 IMU, was mounted to a handheld ultrasound scanner (L7-38, Clarius Mobile Health Corporation, Burnaby, BC, Canada) with a 3D-printed mount. We used two phantom models of the same distal femur, with the patella in two different positions. The first model had the patella in the position of knee flexion, while the second had the patella in the position of knee extension. Our objective is to reconstruct these two models to visualise the different positions of the patella. To ensure that the bone models remained stationary during scanning, we bolted them to a metal plate. The phantoms were encased in a gelatine medium to enable ultrasound scanning. Retro-reflective markers were placed on the bone and the scanner to enable motion capturing. All the devices were connected to a single PC running a Robotic Operation System (ROS) to record all the sensor data and ensure that all the timestamps were recorded from a common clock. 

\subsection{Dataset Collection}
A total of 150 scanning motions were taken, out of which 130 involved the user slowly panning across the phantom while repeating different starting positions and scanning directions. The scanning path was short and contained either 2 or 6 sweeps. The remaining 20 scans were longer and involved the user moving the scanner laterally while panning to cover the entire phantom. In all scans, the user picked up the scanner from a fixed cradle and returned it to the same position at the end.

An OptiTrack motion capture setup was used for ground truth, which has an expected error of less than \SI{1}{\milli\meter}. However, marker occlusions can cause errors in the position estimation. To mitigate this issue, we applied a median filter with a window size of 9 to the camera data, which was recorded at a rate of \SI{200}{\hertz}. The filtered trajectory was used as a reference point for comparing other estimation methods and as training data for the network of the IMU-only method.

\subsection{Tracking methods}
\begin{figure}
	\centering
	\includegraphics[width=0.95\columnwidth]{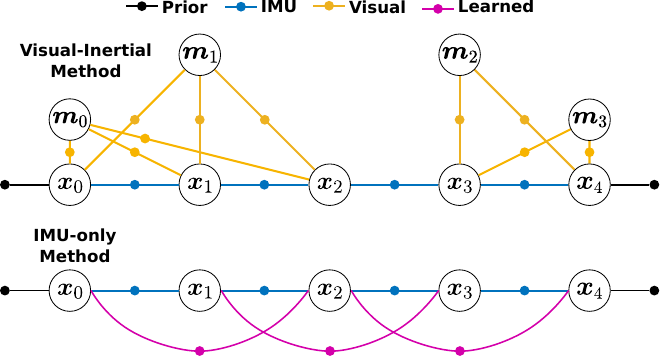}
	\caption{Two example factor graph frameworks for five states $j={0,1,2,3,4}$. On top is the VIO method which uses prior, visual and IMU factors. On the bottom is the learned inertial odometry method which only uses IMU and learned factors. The four types of factors are: prior factors constraining the start and end pose to be equivalent, IMU integration factor from~\cite{Forster2017}, visual landmark factors from \cite{Wisth2022} and our learned displacement factor from~\cite{Buchanan2021CORL}. }
	\label{fig:factor}
	\vspace{-4mm}
\end{figure}

The VIO and IMU-only methods are shown in Fig. \ref{fig:factor} top and bottom, respectively. Both methods were applied to estimate the Maximum-A-Posteriori (MAP) state trajectory of a handheld ultrasound scanner in batch. The state estimator was implemented using the factor-graph library GTSAM \cite{Dellaert2012} and built on the work from \cite{Wisth2022} for VIO, while the neural network for the IMU-only was based on \cite{Buchanan2021CORL} and developed using the PyTorch library.

In this work, we assumed each trajectory has a ``loop closure,'' meaning the scanner trajectory ends with the same position and orientation as at the start. We also assume zero velocity at the start and end (as shown in Fig. \ref{fig:factor}). Two prior factors are used, one for the first pose and one for the last pose. We use standard IMU pre-integration theory to connect consecutive nodes in the graph, as in \cite{Forster2017}. This relates to the pose, velocity, and biases between two consecutive states.

\subsubsection{VIO Method} IMU data was recorded at \SI{200}{\hertz} while the stereo camera was sampled at \SI{30}{\hertz} with one in three frames selected as a keyframe. Therefore, nodes were added to the graph at \SI{10}{\hertz}, with IMU preintegration factors linking each consecutive pair. 

\subsubsection{IMU-only Method} the IMU data samples were buffered in windows of \SI{1}{\second} ($N = 200$). The inference was done at a rate of \SI{30}{\hertz}, which we empirically found to give the best performance. Therefore, each learned displacement links two states \SI{1}{\second} apart and the neural network starts producing outputs after \SI{1}{\second} of data is collected.

The IMU and OptiTrack data collected during the scanning were divided into two sets of 80\% for training and 20\% for validation, respectively. A single model was trained using all data. The Adam optimizer was used with a learning rate of $10^{-6}$ over 2000 epochs and the model minimized validation error. The training process took about 6 hours on an NVIDIA GeForce RTX 3080 Ti GPU. Two scan types were used for both phantoms, labelled as Scan Type A for cross-sections ($x$ direction) and Scan Type B for longitudinal ($y$ direction) sections.

\subsection{Reconstruction}
\label{sec:reconstruction}
The reconstruction was adopted from a previous work \cite{Jia2017} with weighted projective fusion for point cloud integration and surface reconstruction. The final output mesh can be directly compared with ground truth meshes of bone phantom models.

\subsubsection{Surface segmentation}
\begin{figure}
	\centering
	\includegraphics[width=0.7\columnwidth]{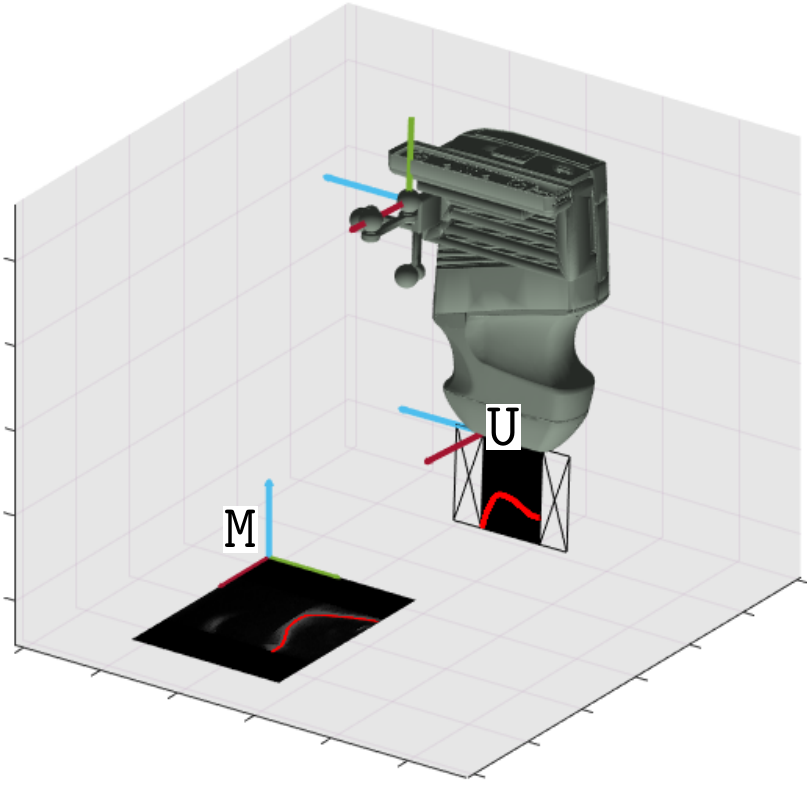}
	\caption{Ultrasound images collected through the Clarius API are in a size of $640 \times480$. The origin of the ultrasound image frame $\mathtt{M}$ was defined as pixel (0,0). A fixed, known transformation from $\mathtt{M}$ into the ultrasound scanner's frame $\mathtt{U}$ was applied to all segmented pixels to project them into 3D points. Finally, the 3D points in $\mathtt{U}$ were transformed into the global coordinate system using the selected odometry source.}
	\label{fig:calibration}
	\vspace{-4mm}
\end{figure}

The bone surface was segmented from ultrasound images using the method in \cite{Jia2016}. The segmented pixels are then projected into 3D as points (as shown in Fig. \ref{fig:calibration}). With the trajectory estimate from the tracking module, the projected points are accumulated into point clouds according to the ``sweep'' number over the bone. Finally, the reconstructed point clouds are reviewed and cleaned by removing mislabelled points. Several point clouds of the bone surface corresponding to a separate sweep were carried on for further evaluation.

\subsubsection{Alignment of Point Cloud Sweeps}
For VIO and IMU-only methods there is incremental drift while estimating the scanner's pose. Knowing that we are scanning the same bone over several sweeps, we can correct this drift and align the sweep points using  Iterative Closest Point (ICP)\cite{Pomerleau2015} with an open-source library \textit{libpointmatcher}\cite{Pomerleau2013}. ICP was constrained to position and yaw (rotation about the gravity axis) only, considering the acceleration, pitch and roll estimate from the IMU were highly accurate over a short time period of a single scan. Although errors caused by tracking drift while moving across the bone cannot be corrected as the sweeps are treated as rigid point clouds, we assume that the drift incurred over this small distance is not enough to severely degrade the reconstruction.

While the Optitrack-estimated sensor pose is not subject to drift, aligning point clouds also slightly improves reconstruction for this configuration.

\subsubsection{Surface Reconstruction and Meshing}
\label{sec:surface-reconstruction}
The open source library Open3D\cite{Zhou2018} were used to perform Truncated Signed Distance Fields (TSDF) representation and integration algorithm\cite{Newcombe2011} to fuse the aligned point clouds into a smooth mesh surface. TSDFs are a volumetric representation where voxel values correspond to the signed distance from the surface boundary, being positive outside the surface and negative when penetrating inside the surface. TSDF integration was computed with \SI{2}{\milli\meter} resolution and \SI{1}{\centi\meter} truncation with \SI{1}{\milli\meter} voxel filtering and Point-to-Plane matching to improve robustness to outlier points.

Alignment and surface reconstruction were done in a loop whereby point clouds were first aligned and then integrated into the surface. A point cloud sampling of the TSDF surface is then extracted and used to align the next point cloud. Once all the sweeps have been aligned and integrated into the final surface, then the point cloud is converted to a mesh using the Poisson method\cite{Kazhdan2006}.

\section{Results and Discussion}

\subsection{Tracking Results}
Table \ref{tab:estimation-results} shows the level of estimation error of each tracking method compared to the Optitrack trajectory, which was considered the ground truth. Visual-inertial and deep learning inertial values are presented with and without the loop closure constraint from the start and end of prior factors.

Both the Absolute Pose Error (APE) and Relative Pose Error (RPE) translation components \cite{Kummerle2009} were computed. APE gives the average error of each trajectory compared to the ground truth trajectory at any point in time. RPE, on the other hand, shows the rate of drift over a given distance. As shown in Table \ref{tab:estimation-results}, RPE was computed over \SI{20}{\milli\meter} segments, indicating that the value represents the average drift for every \SI{20}{\milli\meter} travelled. 

Adding the loop closure slightly decreases the average position error (APE) in most cases except for Bone 2 Scan A, but it has no effect on the RPE. This is because the VIO method is already highly accurate and drifts an average of \SI{1.8}{\milli\meter} per \SI{20}{\milli\meter}. In contrast, the IMU-only methods have larger errors than VIO. Even though adding the loop closure does reduce both RPE and APE for most trajectories, it remains higher than VIO.

\begin{figure}
	\centering
	\includegraphics[width=0.95\columnwidth]{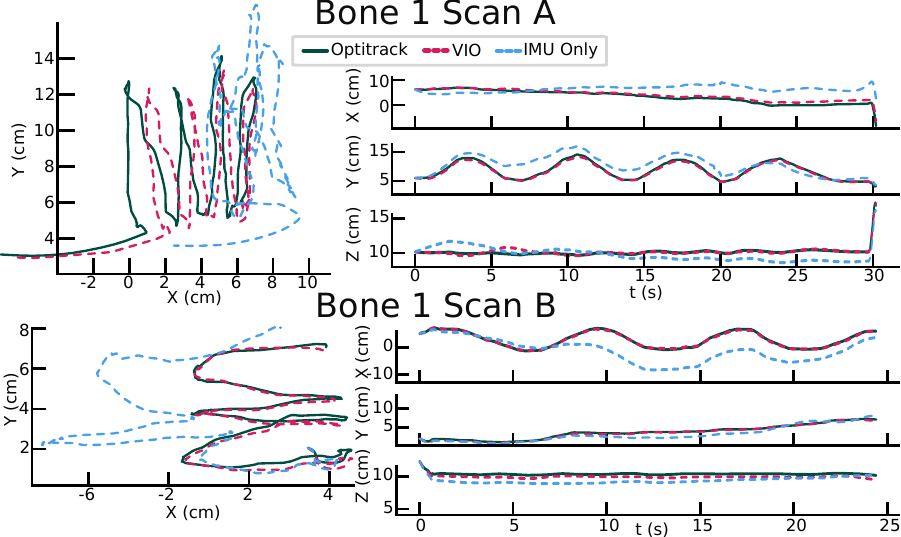}
	\caption{Trajectories for the different algorithms while scanning Bone 1. Visual-inertial and learned inertial methods both use the loop closure constraint. In these plots, the trajectory is cropped to start when the scanner is in contact with the bone and cropped to end when scanning is finished. The first pose is aligned with ground truth in position and yaw, not roll or pitch.}
	\label{fig:bone1-trajectory}
 \vspace{-4mm}
\end{figure}

\begin{figure}
	\centering
	\includegraphics[width=0.95\columnwidth]{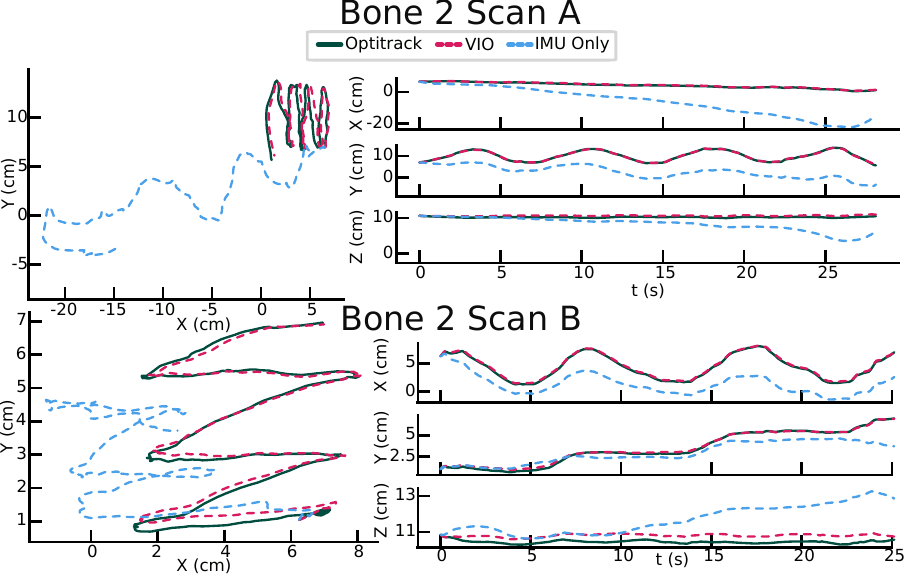}
	\caption{Trajectories for the different algorithms while scanning Bone 2. Visual-inertial and learned inertial methods both use the loop closure constraint. Alignment is the same as for Fig.~\ref{fig:bone1-trajectory}.}
	\label{fig:bone2-trajectory}
  \vspace{-3mm}
\end{figure}

The trajectories for all six scans for Bone 1 (Fig. \ref{fig:bone1-trajectory}) and Bone 2 (Fig. \ref{fig:bone2-trajectory}) were plotted and showed a clear difference in the magnitude of error between VIO and IMU. The VIO methods were almost perfectly aligned with Optitrack, whereas the IMU-Only method had more significant errors. However, for five of the six trajectories, the method largely followed the Optitrack trajectory with a slight offset. It is evident that the network has learned the scanning motion and is providing useful information to the estimator. In the reconstruction step, we corrected these errors using ICP.

\begin{table}[t!]
	\centering
	\caption{Position Estimation Error [mm]}
		\resizebox{0.95\columnwidth}{!}{
		\begin{tabular}{l|cc|cc|cc|cc}  
			\toprule
			\textbf{Method} & \multicolumn{4}{c|}{\textbf{APE}}& \multicolumn{4}{c}{\textbf{RPE (\SI{20}{\milli\meter})}}\\
			\midrule
			\textbf{Bone} & \multicolumn{2}{c|}{Bone 1} & \multicolumn{2}{c|}{Bone 2} 
			& \multicolumn{2}{c|}{Bone 1} & \multicolumn{2}{c}{Bone 2} \\
			\midrule
			\textbf{Scan Type} & A  & B & A  & B & A & B & A  & B \\
			\midrule
			VIO & 8.7  & 3.8 & 3.5 & 4.2 & 2.3  & 1.9 & 1.6 & 1.2 \\
			VIO - Loop &  8.3  & 3.8 & 4.9 &  4.2 & 2.3  & 1.9 & 1.6 & 1.2 \\
			IMU-Only &   53.2  & 124.4 & 112.2 &  18.9 & 10  & 14.7 & 14.6 & 9.7 \\
			IMU-Only - Loop & 49.0 & 41.9 & 132.7 & 34.7 & 9.9  & 11.8 & 15.3 & 9.7 \\
			\bottomrule
		\end{tabular}
			}
	\label{tab:estimation-results}
 \vspace{-4mm}
\end{table}

\subsection{Reconstruction Results}
Table \ref{tab:reconstruction-results} shows the mean and worst distance errors of each method when compared to the ground truth mesh model of the phantom bone. The motion capture method and the visual-inertial method had average errors of \SI{1.21}{\milli\meter} and \SI{1.25}{\milli\meter} respectively, with only a 3\% difference between them. However, the IMU-only method had an average error of \SI{1.85}{\milli\meter}, which was 35\% higher than the motion capture method. This is because the IMU-only method has higher drift rates

\begin{figure}
	\centering
	\includegraphics[width=0.95\columnwidth]{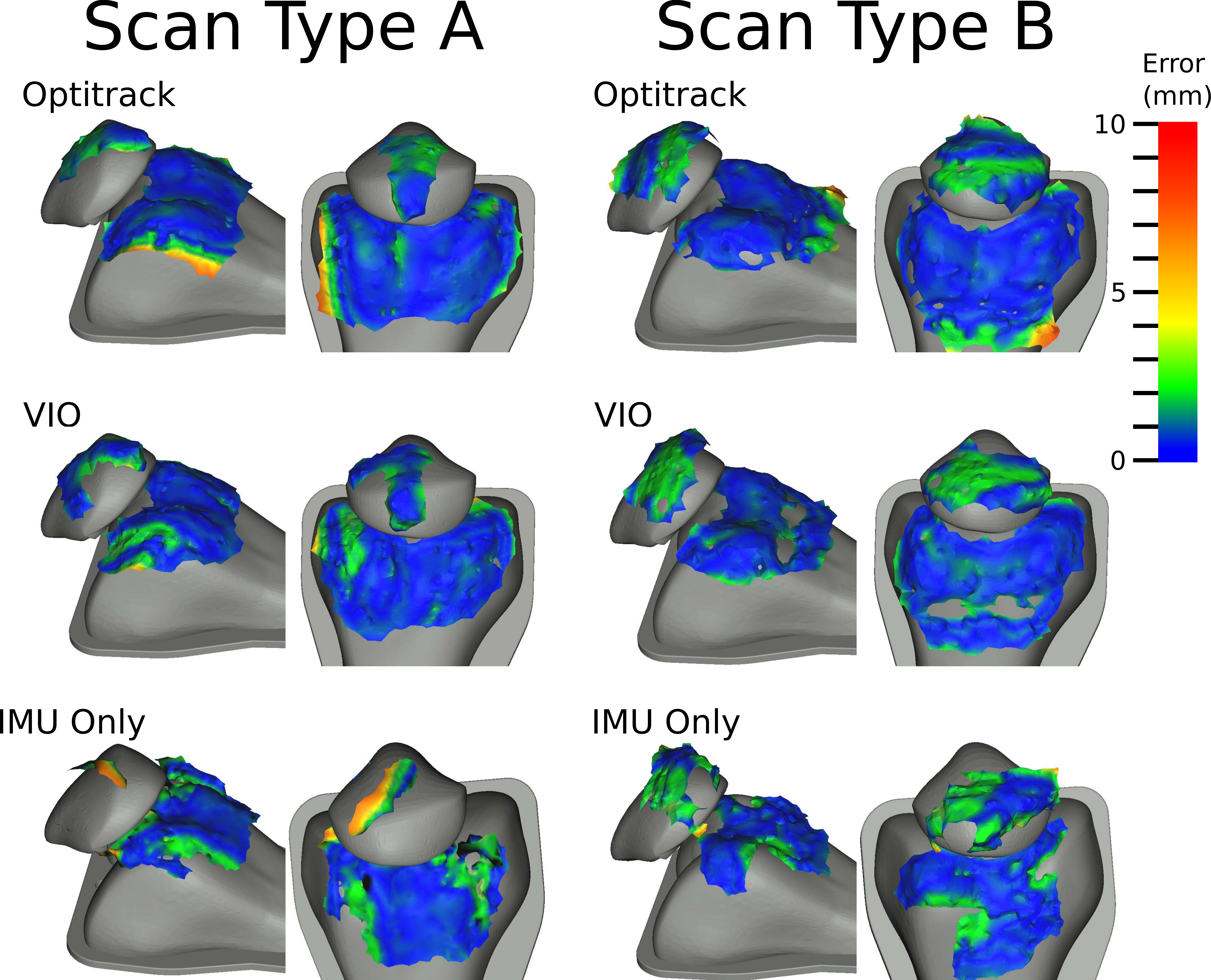}
	\caption{Reconstructions of Bone 1 which is in 40$\degree$ flexion. We show results for each estimation method and for both scanning types A and B.}
	\label{fig:B1-reconstruction}
\end{figure}

\begin{figure}
	\centering
	\includegraphics[width=0.95\columnwidth]{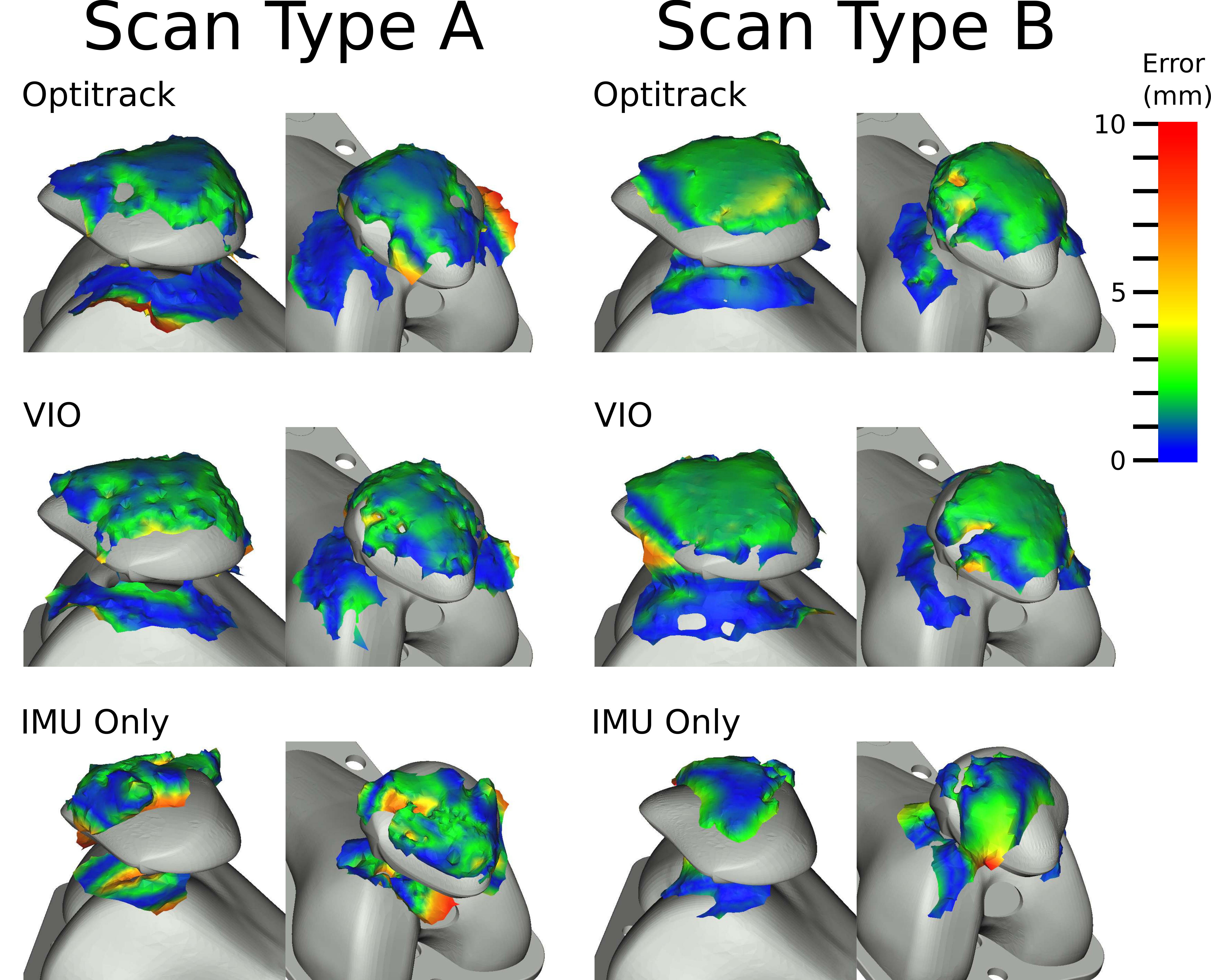}
	\caption{Reconstructions of Bone 2 which is in full extension. The patella is now directly over the distal femur. We show results for each estimation methods and for both scanning types A and B.}
	\label{fig:B2-reconstruction}
\end{figure}

Fig. \ref{fig:B1-reconstruction} and Fig. \ref{fig:B2-reconstruction} present the reconstructed models of Bone 1 and Bone 2 for each scanning type. The colourised mesh represents the distance error from the ground truth model, and the surface reconstruction is overlaid on top of the ground truth model for each bone. It is important to note that the mesh reconstructions may contain some errors, such as a flaring out of the mesh at the edges, which is observable in Bone 1 Scan A with OptiTrack (Fig. \ref{fig:B1-reconstruction}).

In this study, we demonstrated a slightly higher error rate compared to other studies, without requiring expensive external infrastructure. Reference \cite{Mahfouz2021} used EM trackers to track the scanner's movement on the skin and reported an average error of \SI{1.05}{\milli\meter} while scanning the distal femur and proximal tibia of 15 cadaveric specimens. Reference \cite{Kerr2017} used a robotic arm to perform the scanning, and they reported an average error of \SI{0.85}{\milli\meter} across two distal femurs (one human model and one bovine) using their best results, with an average worst error of \SI{9.06}{\milli\meter}.

Other reconstruction methods can cover a wider area of the bone, but they may introduce additional problems. For example, using a statistical shape model like in \cite{Mahfouz2021} can lead to biased reconstructions. Similarly, submerging the sensor in water as in \cite{Kerr2017} is not a practical solution. In contrast, our method involves using a general-purpose scanner held by a human and does not require an SSM. This approach makes our method more practical and versatile.

\begin{table}
	\centering
	\caption{Mean/Worst Reconstruction Error [mm]}
		\resizebox{0.95\columnwidth}{!}{
		\begin{tabular}{l|cc|cc}  
			\toprule
			\textbf{Method} & \multicolumn{2}{c|}{\textbf{Bone 1}} & \multicolumn{2}{c}{\textbf{Bone 2}} \\
			\midrule
			\textbf{Sweep Type} & A & B & A & B \\
			\midrule
			Optitrack &  0.99~/~7.71  & 1.00~/~7.50 & 1.53~/~9.95   & 1.76~/~5.11 \\
			VIO &  0.85~/~5.2    & 1.10~/~3.93 & 1.70~/~6.43  &  1.80~/~6.32 \\
			IMU-Only &  1.8~/~7.4   & 1.51~/~7.67 & 2.56~/~10.03 & 1.98~/~7.93 \\
			\bottomrule
		\end{tabular}
			}
	\label{tab:reconstruction-results}
 \vspace{-4mm}
\end{table}

\subsection{Future Work}
Our work shows that using outward-facing VIO to track the pose of an ultrasound probe is accurate enough to perform markerless, handheld ultrasound reconstruction with ICP-based reconstruction. However, the IMU-only method is not currently practical due to its high drift rate. Although incorporating more prior information and accumulating point clouds over shorter distances may help, it could make ICP matching more challenging. Additionally, introducing an SSM could result in bias in the reconstruction, but it may be useful for estimating the joint position, provided that there are no bone abnormalities. To determine the impact of available data on the results, we tested models with less data and found that RPE only increased by 2.3\% when half the data was used. We have also discovered that having more data does not significantly enhance the IMU-only tracking. Instead, we recommend exploring new neural network architectures.

\section{Conclusion}
We introduced a new method for creating a 3D reconstruction of bones that doesn't require the use of markers or external infrastructure. The technique involves using a handheld ultrasound scanner and combining it with recent robotic estimation techniques such as visual-inertial odometry and learning inertial motion tracking to determine the position of the scanner. Our experiment indicates that using outward-facing VIO to generate bone surfaces results in errors that are comparable to those obtained using external tracking infrastructure. Precise and streamlined 3D freehand ultrasound possesses significant potential across various clinical applications, for example, the method described in the current study can be used to visualise and measure the position of the patella with varying degrees of knee flexion in patients with painful TKA.

\bibliographystyle{IEEEtran.bst}
\bibliography{references.bib}
\end{document}